\documentclass[prb,aps,twocolumn,showpacs,preprintnumbers,amsmath,amssymb]{revtex4-1}
\usepackage{graphicx}
\usepackage{amsmath}
\begin{document}
\title{First-principles computational study of defect clustering in solid solutions of ThO$_{2}$ with trivalent oxides}

\author{Vitaly Alexandrov}
\affiliation{Department of Chemical Engineering and Materials Science and NEAT ORU, University of California, Davis, California 95616, USA \\
Department of Materials Science and Engineering, University of California, Berkeley, California 94720, USA} 
\author{Niels Gr{\o}nbech-Jensen}
\affiliation{Department of Applied Science, University of California, Davis, California 95616, USA}
\author{Alexandra Navrotsky}
\affiliation{Peter A. Rock Thermochemistry Laboratory and NEAT ORU, University of California, Davis, California 95616, USA \\
Department of Chemical Engineering and Materials Science, University of California, Davis, California 95616, USA} 
\author{Mark Asta}
\affiliation{Department of Materials Science and Engineering, University of California, Berkeley, California 94720, USA \\
Department of Chemical Engineering and Materials Science and NEAT ORU, University of California, Davis, California 95616, USA}

\date{\today}

\begin{abstract} 
The energetics of mixing and defect ordering in solid solutions of fluorite-structured ThO$_{2}$ with oxides of trivalent cations (Sc, In, Y, Nd, La)
are investigated by electronic density-functional-theory (DFT). Through DFT calculations of structures enumerated by lattice-algebra techniques, we 
identify the lowest-energy patterns for defect clustering for four separate dopant concentrations. The most stable structures are characterized by a 
repulsive interaction between nearest-neighbor vacancies on the oxygen sublattice. The enthalpies of formation with respect to constituent oxides 
are positive for all dopants considered, and show a tendency to decrease in magnitude as the size and electronegativity of the trivalent dopant
decrease. Due to the small positive formation enthalpies and low oxygen-vacancy binding energy with La dopants, La$_{2}$O$_{3}$-ThO$_{2}$ 
solid solutions are predicted to have relatively high ionic conductivities relative to those for the other aliovalent dopants considered. Our results 
are compared with those for the more widely studied ZrO$_{2}$ and CeO$_{2}$ fluorite-structured solid solutions with trivalent cations.

\end{abstract}
\maketitle

\section{Introduction}
Design of solid-state oxygen ion conductors, oxygen sensors and separation membranes\cite{Minh1993,Kharton2000,Steele2001,Jiang2004,Kharton2008} 
typically involves aliovalent cation doping of metal oxides to form charge-compensating oxygen vacancies. The observation of high oxygen 
conductivity in aliovalently doped fluorite-structured oxides has motivated extensive experimental and theoretical research aimed at relating the
structural and chemical defect properties in these systems to their ionic conductivities.  Recent studies of trivalent-doped zirconia, hafnia, 
ceria and thoria \cite{Trubelja1991,Goff1999,Lee2003,Chen2006,Navrotsky2007,Avila-Paredes2009,Aizenshtein2010} have revealed a strong correlation between oxygen 
conductivity and the mixing thermodynamic properties.  Specifically, enthalpies of formation ($\Delta H^{f}$) derived from calorimetric measurements
tend to become more positive with dopant concentration ($x$) at low solute concentrations, in a manner consistent with a regular solution behavior; near some maximum composition ($x_{max}$), however, $\Delta H^{f}$ is observed to level off or become less positive with further doping.\cite{Chen2006,Lee2003,Aizenshtein2010} 
The value of $x_{max}$ is found to coincide with the maximum in measured ionic conductivity.\cite{Avila-Paredes2009}  These experimental 
results suggest that the mobile oxygen vacancies begin to strongly associate and form stable, relativley immobile, clusters for dopant concentrations near and beyond x$_{max}$.\cite{Goff1999,Chen2006,Navrotsky2007,Lee2003,Tien1963} However, the detailed nature of these defect clusters, and how they vary 
in different systems, remains incompletely understood.

In the technologically important yttria-stabilized zirconia (YSZ) system, atomistic simulations
\cite{Khan1998,Zacate2000,Bogicevic2001a,Bogicevic2001,Predith2008,Xia2009} agree with neutron and X-ray diffraction experiments\cite{Ray1980,Goff1999} 
that oxygen vacancies tend to align along the $<$111$>$ direction as third nearest neighbors on the oxygen sublattice, and prefer to associate with 
the smaller host Zr ions rather than with the larger Y dopants.  Further, DFT studies\cite{Bogicevic2001a,Bogicevic2001,Predith2008} have shown that 
these defect-clustering tendencies result in the stability of two ordered ground-state structures:  the experimentally observed 
$\delta$-Zr$_{3}$Y$_{4}$O$_{12}$ compound and an as-yet unobserved ZrO$_{2}$-rich phase. Through studies of other trivalent and divalent dopants in 
ZrO$_{2}$, it has been shown that the formation enthalpies of doped ZrO$_{2}$ compounds are most positive for small dopant ions. The results also 
demonstrate a tendency for oxygen vacancies to associate with the smaller cation species (Zr or dopant) in a given system. In classical atomistic 
simulations of the energetics of defect cluster formation in trivalent-doped CeO$_{2}$ (Ref.\cite{Pryde1995,Minervini1999}) it has been demonstrated 
that the atomic geometry of defect clusters is controlled by a combination of electrostatic and elastic energy effects, the latter being dependent 
on the dopant ion size. The binding energies between dopant cation and oxygen vacancies also have been shown to be a strong function of the dopant 
ion size, with the smallest dopants exhibiting greatest binding energies with nearest-neighbor oxygen vacancies.

Solid solutions of ThO$_{2}$ with aliovalent trivalent cations also can be utilized for solid electrolyte applications, similar to doped zirconia and 
ceria.  ThO$_{2}$ also has been gaining increasing attention due to the renewed interest in the Th--U fuel cycle which offers a number of advantages 
over the U--Pu cycle including: higher thermal conductivity of ThO$_{2}$ relative to UO$_{2}$ based fuels and lower concentrations of long-lived transuranic elements.\cite{Lung1998,Lombardi2008}  The trivalent dopants Y, Nd and La are important products in the thorium-based nuclear 
cycle.  Recently, Y$_{2}$O$_{3}$- and La$_{2}$O$_{3}$-doped ThO$_{2}$ have been the subject of calorimetric studies of mixing energetics.
\cite{Aizenshtein2010} As in doped CeO$_{2}$ systems, the oxygen-ion conductivity in Y$_{2}$O$_{3}$-doped ThO$_{2}$ shows the 
aforementioned correlation between the maximum in formation enthalpy and oxygen conductivity.

The key to understanding the thermodynamics of defect clustering and its correlation with conductivity is to identify underlying interatomic 
interactions between dopants and oxygen vacancies. In solid solutions of aliovalently doped oxides, electrostatic interactions between ions with 
different charges and strain-mediated interactions due to the size mismatch between host and dopant atoms, and the presence of oxygen vacancies, are 
general in nature. Thus, if these are the primary factors driving defect association, the nature of the defect clusters might be expected to be generic 
across fluorite-structured systems, depending only on the charge and size of the aliovalent dopant relative to the host. The analysis of 
driving forces for defect clustering can be more complex if other factors play a significant role. For example, in the case of YSZ, it has been 
suggested\cite{Bogicevic2001a,Stefanovich1994,Stapper1999} that the preference of Zr cations to be seven-fold coordinated by oxygen in the 
lowest-energy monoclinic phase of ZrO$_{2}$ leads to a driving force for association with oxygen vacancies in the eight-fold coordinated fluorite 
structure. 

ThO$_{2}$ is an interesting system from the standpoint of studying defect interactions in fluorite-structured compounds, as (in contrast to ZrO$_{2}$) 
it is a system that displays a stable fluorite structure in pure form, and (in contrast to CeO$_{2}$) it displays a low equilibrium solubility of oxygen 
vacancies and no tendency toward reduction. In this study, we exploit first-principles DFT-based calculations combined with the lattice-algebra structure enumeration technique to study 
mixtures of ThO$_{2}$ with trivalent dopants.  To investigate the energetics and the nature of defect clustering, we examine a set of dopants
spanning a range of ionic sizes and electronegativities, as shown in Table \ref{table1}.

\begin{table}
\caption{\label{table1}The effective ionic radii of Shannon\cite{Shannon1976} $r$ (\AA) for trivalent dopants in the octahedral coordination\footnote{The ionic radius of the host Th$^{4+}$ is 0.94 \AA \ and its Pauling electronegativity is 1.3.} and Pauling electronegativities $\chi$.}
\begin{ruledtabular}
\begin{tabular}{cccccc}
          & Sc & In & Y & Nd & La \\
     $r$    &  0.75 & 0.80 & 0.90 &  0.98 & 1.03 \\
     $\chi$  &  1.36 & 1.78 & 1.22 & 1.14 & 1.10 \\
        \end{tabular}
\end{ruledtabular}
\end {table}

\section{Computational methodology}

First-principles calculations were carried out within the DFT plane wave formalism using the GGA-PBE exchange-correlation functional\cite{Perdew1996} and 
the projector augmented wave (PAW) potentials\cite{Blochl1994} as implemented in the Vienna \textit{Ab initio} Simulation Package 
(VASP).\cite{Kresse1993,Kresse1994} It has been demonstrated\cite{Sevik2009} that the inclusion of Hubbard-type on-site electron interaction within the DFT + $U$ approach\cite{Dudarev1998} for calculation of structural, elastic and electronic properties  of ThO$_{2}$ changes the results only marginally as compared with pure DFT calculations. We employ the potentials labeled Th, O, Sc\_sv, In\_d, Y\_sv, Nd and La in the VASP PBE library; for  Sc\_sv, Y\_sv 
and In\_d these potentials include semicore states as valence electrons.  In all calculations we employ a cutoff energy of 500 eV which is 25\% higher 
than the default value for the potentials chosen to ensure convergence of the calculated stress tensors for volume relaxations.  All atomic structure relaxations were performed starting from configurations of atoms in ideal fluorite 
positions and optimizing cell volume, cell shape and all ionic positions within a cell by applying a conjugate-gradient algorithm until atomic forces
were converged to a magnitude less than 0.1 meV/\AA. In these relaxation runs, we employed the Monkhorst-Pack scheme\cite{Monkhorst1976} to sample the 
Brillouin zone, using meshes with a density of at least 4$\times$4$\times$4.  Subsequent to the structural optimizations, an additional static run for
each final optimized geometry was made using accurate tetrahedron integration with Bl\"{o}chl corrections,\cite{Blochl1994a} with a sufficiently dense sampling of the 
Brillouin zone to converge the total energy to better than 10$^{-6}$ eV. For structures involving the trivalent Nd dopant, which contains three valence 
$f$ electrons, calculations were performed spin-polarized, using a ferromagnetic arrangement of the local magnetic moments (for a few structures both 
ferromagnetic and antiferromagnetic ordering was considered, and the former was found to be lower in energy).

\begin{figure}
\includegraphics[scale=0.7]{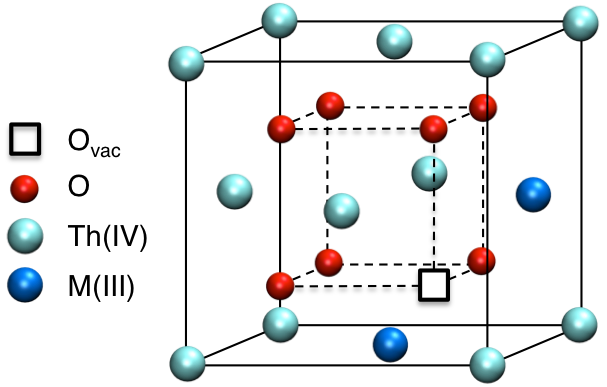}\\
\caption{\label{fig1}ThO$_{2}$ cubic fluorite structure with two trivalent dopants on the cation $fcc$ sublattice
and one charge-compensating oxygen vacancy on an anion simple cubic sublattice.}
\end{figure}

In fluorite-structured ThO$_{2}$ doped by oxides of trivalent cations, one has to account for binary disorder on both cation (Th$^{4+}$, M$^{3+}$, where 
M denotes the trivalent dopant species) and anion (O$^{2-}$, O$_{vac}$) sublattices (see Fig. \ref{fig1}). In order to probe the configurational space and 
sample different possible atomic arrangements for given dopant concentrations, we employed a structure enumeration technique making use of the 
multicomponent multisublattice formalism previously developed to study alloys and recently applied to YSZ.\cite{Bogicevic2001a,Bogicevic2001,Predith2008} 
We performed a complete structure enumeration in supercells consisting of up to nine primitive $fcc$ unit cells (for the present study we focus on compositions
where we enumerated structures only up to six unit cells) using the Alloy Theoretic Automatic Toolkit (ATAT).\cite{Walle2002,Walle2009} These enumerated 
structures were considered for dopant concentrations of $x$=0.33 and 0.5 cation mole fraction.  We also employed supercells, composed of 
2$\times$2$\times$2 conventional fluorite unit cells (96 atoms), to study defect energies at the more dilute compositions of $x$=0.0625 and 0.125, 
corresponding respectively to a single vacancy with two dopant cations, and two vacancies with four dopants.

Formation enthalpies of solid solutions normalized per cation can be expressed with respect to the constituent oxides as
$$\Delta H^{f} = E[Th_{1-x}M_{x}O_{2-0.5x}] - (1-x)E[ThO_{2}] - xE[MO_{1.5}],$$
where $E[ThO_{2}]$ denotes the energy (per cation) of fluorite-structured ThO$_{2}$.  Similarly, $E[MO_{1.5}]$ denotes the energy (per cation) of 
Sc$_{2}$O$_{3}$, In$_{2}$O$_{3}$ and Y$_{2}$O$_{3}$ in the experimentally-observed cubic bixbyite (C-type sesquioxide) structure, or Nd$_{2}$O$_{3}$ and
La$_{2}$O$_{3}$ in the observed low-temperature hexagonal A-type structure.  The energy per cation for a solid solution with a cation dopant concentration 
$x$ is denoted as $E[Th_{1-x}M_xO_{2-x/2}]$.  With the above definition, positive values of $\Delta H^f$ correspond to an energetic tendency for phase 
separation of the solid solutions into the respective constituent oxides.

\section{Results}
ThO$_{2}$ compounds form in the cubic fluorite structure in which the cations form an $fcc$ sublattice and anions form a simple cubic sublattice 
(Fig.~\ref{fig1}). Doping ThO$_{2}$ by trivalent cations introduces one charge-compensating oxygen vacancy per two trivalent dopants and leads to a coupled 
intra- and intersublattice interactions between all the species present (Th$^{4+}$, M$^{3+}$, O$^{2-}$, O$_{vac}$). To investigate the nature of these 
interactions, and the geometries of the defect clusters that they favor energetically, we investigate the formation energetics of a variety of 
superstructures for relatively dilute and concentrated compositions in sub-sections A and B, respectively.

\subsection{Defect association in the dilute regime}
In this section, we focus on systems with relatively dilute dopant concentrations employing a 2$\times$2$\times$2 supercell of the host ThO$_{2}$ crystal 
with one and two electrically neutral M--O$_{vac}$--M units, corresponding to $x$=6.25\% and 12.5\% dopant concentrations, respectively. We focus on a 
comparison of the energetics for under-sized Sc, intermediately-sized Y  and over-sized La dopants (Table \ref{table1}). We examine a number of plausible 
defect arrangements and obtain only positive formation enthalpies in agreement with recent calorimetric data for Y and La dopants.\cite{Aizenshtein2010} 

\begin{table}
\caption{\label{table2}Formation enthalpies (kJ/mol-cation) of the most stable defect associates for Th$_{1-x}$M$_x$O$_{2-0.5x}$ (M = Sc, Y, La) systems, 
as derived from calculations employing 2$\times$2$\times$2 supercells of the host ThO$_2$ conventional unit cell.  The stable defect clusters correspond 
to a nearest-neighbor M--O$_{vac}$--M triangular cluster.}
\begin{ruledtabular}
\begin{tabular}{cccc}
                   $x$ & Sc & Y & La \\
                   \hline
                    & \multicolumn{3}{c}{Formation enthalpy} \\
                   \cline{2-4} \\
 0.125  & 14.4 & 6.3 & 3.4 \\
0.0625 & 8.8 & 3.4 & 2.0 \\
                    & \multicolumn{3}{c}{Binding energy} \\
                    \cline{2-4} \\
0.0625 & -1.9 (-1.9) & -1.1 (-1.1) & -0.2 (-0.6) \\                  
        \end{tabular}
\end{ruledtabular}
\end{table}

For the most dilute dopant concentration (6.25\%) the most energetically stable clusters correspond to the compact triangle geometry involving M and 
O$_{vac}$ as nearest-neighbor pairs.  The calculated formation energies of these compact clusters are listed in Table \ref{table2}, and are seen to 
significantly decrease in magnitude from Sc to Y to La. We also study the energetics of clustering between two triplet defect clusters (12.5\% defect 
level) by considering a number of different atomic arrangements. We find that the lowest-energy arrangements correspond to two nearest-neighbor triplet 
M--O$_{vac}$--M clusters, with the two oxygen vacancies aligned as second neighbors on the oxygen sublattice, i.e., along the $<$110$>$ direction, with 
three dopant cations and one host thorium ion composing tetrahedron around the oxygen vacancy. The formation energies of these clusters are also listed 
in Table \ref{table2} and again show a trend of decreasing magnitudes for $\Delta H^f$ going from Sc to Y to La.

We further estimate the binding energies between the trivalent dopants and oxygen vacancies for the most stable defect clusters. The binding energy here is defined as the difference in energy between a system with a single defect cluster and one with isolated defects, such that negative values
indicate an energetically stable defect cluster.  Due to the relatively small supercells used in this study, we cannot derive highly accurate values
for these binding energies.  Rather we explore trends by estimating the energy of the isolated defects as the energy of the supercell in which the 
defects are separated as far as possible within the 2$\times$2$\times$2 supercell.

In the case of the M--O$_{vac}$--M triplet cluster, the binding energy can be split into two parts corresponding to subsequent breaking of the first 
and the second nearest-neighbor M--O$_{vas}$ bond. This allows one to estimate whether or not the dopant-vacancy interactions are 
pairwise-additive in nature.  The energy to break the first and second nearest-neighbor M--O$_{vac}$ bonds is given in the final row of Table 
\ref{table2}.  For Sc and Y the interactions are found to be relatively short-range in nature, and the binding energies presented in Table 
\ref{table2} do indicate nearly pairwise-additive interactions between the dopant and oxygen vacancy (i.e., the energy to break the first and 
second nearest-neighbor M--O$_{vac}$ bond is identical within the precision of the DFT calculations). In the case of La, however, the 
interactions are found to be much longer ranged; the interaction energy between a La ion and a La--O$_{vac}$ pair is found to vary appreciably with 
distance out to the largest separation attainable in the 2$\times$2$\times$2 supercell. For the La dopant, it can be seen that the interactions 
between dopant and oxygen vacancy are not pairwise additive, as it requires nearly three times the energy to break the second La--O$_{vac}$ bond as 
it does the first. The longer-range an non-pairwise nature of the interactions between La and O$_{vac}$ is likely a manifestation of 
elastically-mediated interactions, as supercell calculations performed by constraining the ions to ideal fluorite lattice sites yielded interaction 
ranges and pairwise-additive behavior very similar to that described above for the Y dopants. The fact that elastic interactions can lead to 
long-range and non-pairwise interactions has been demonstrated by Bugaev et al.\cite{Bugaev2002}  It is nevertheless important to point out that 
the binding energies are particularly small for La system. More generally, we note that the results in Table \ref{table2} show a trend towards 
decreasing magnitude of the binding energies going from Sc to Y to La.

\subsection{Defect association in the concentrated regime}

\begin{figure}
\includegraphics[scale=0.7]{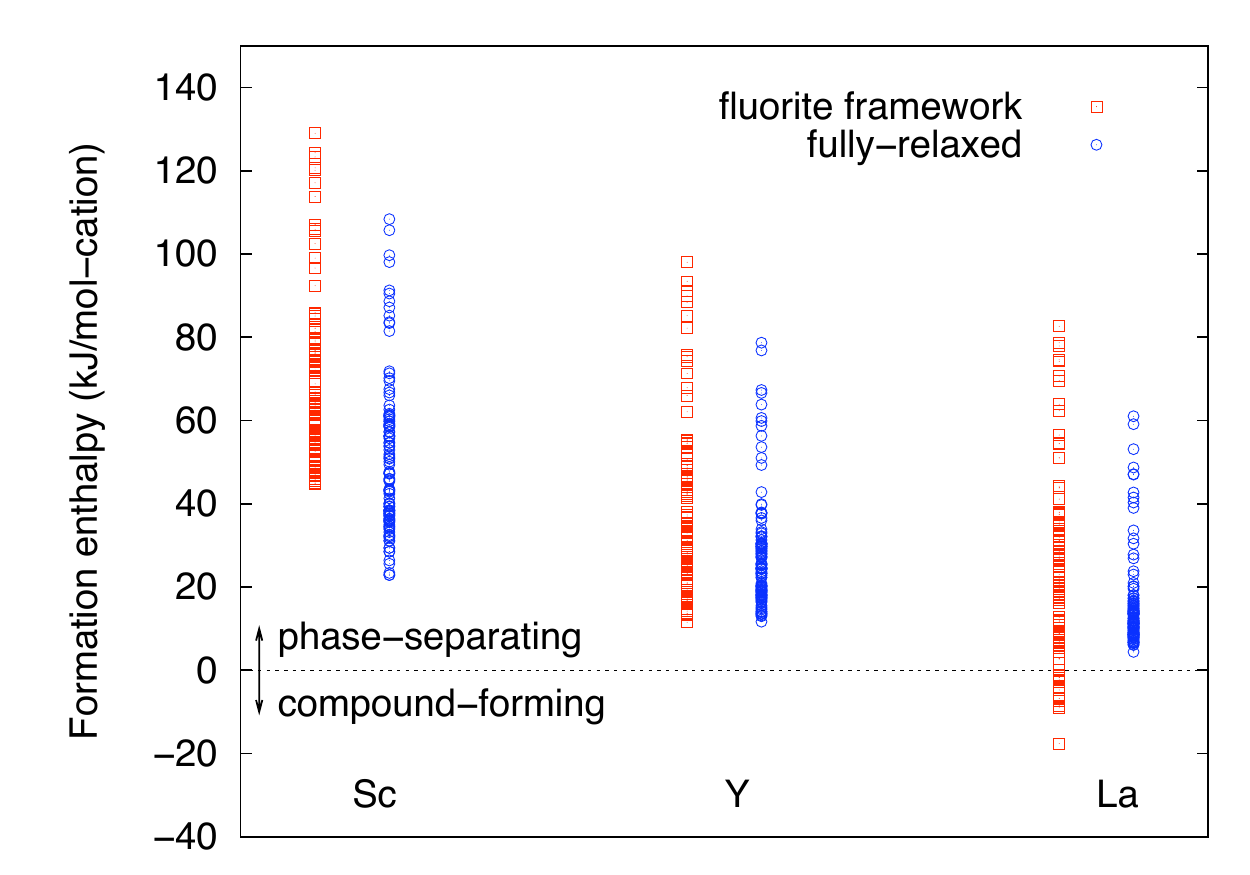}\\
\caption{\label{fig2}Formation enthalpies (kJ/mol-cation) of Th$_{4}$M$_{2}$O$_{11}$ (M = Sc,  Y and La) solid solutions corresponding to $x$=0.33 doping 
level. Red boxes represent results obtained considering only volume relaxations, with the ionic positions for the end-member compounds and mixtures fixed 
at ideal fluorite lattice sites.  The blue circles represent results obtained for fully-relaxed structures, and with end-member M$_2$O$_3$ compounds in 
their experimentally-observed low-temperature structure.}
\end{figure}

\begin{figure}
\includegraphics[scale=0.7]{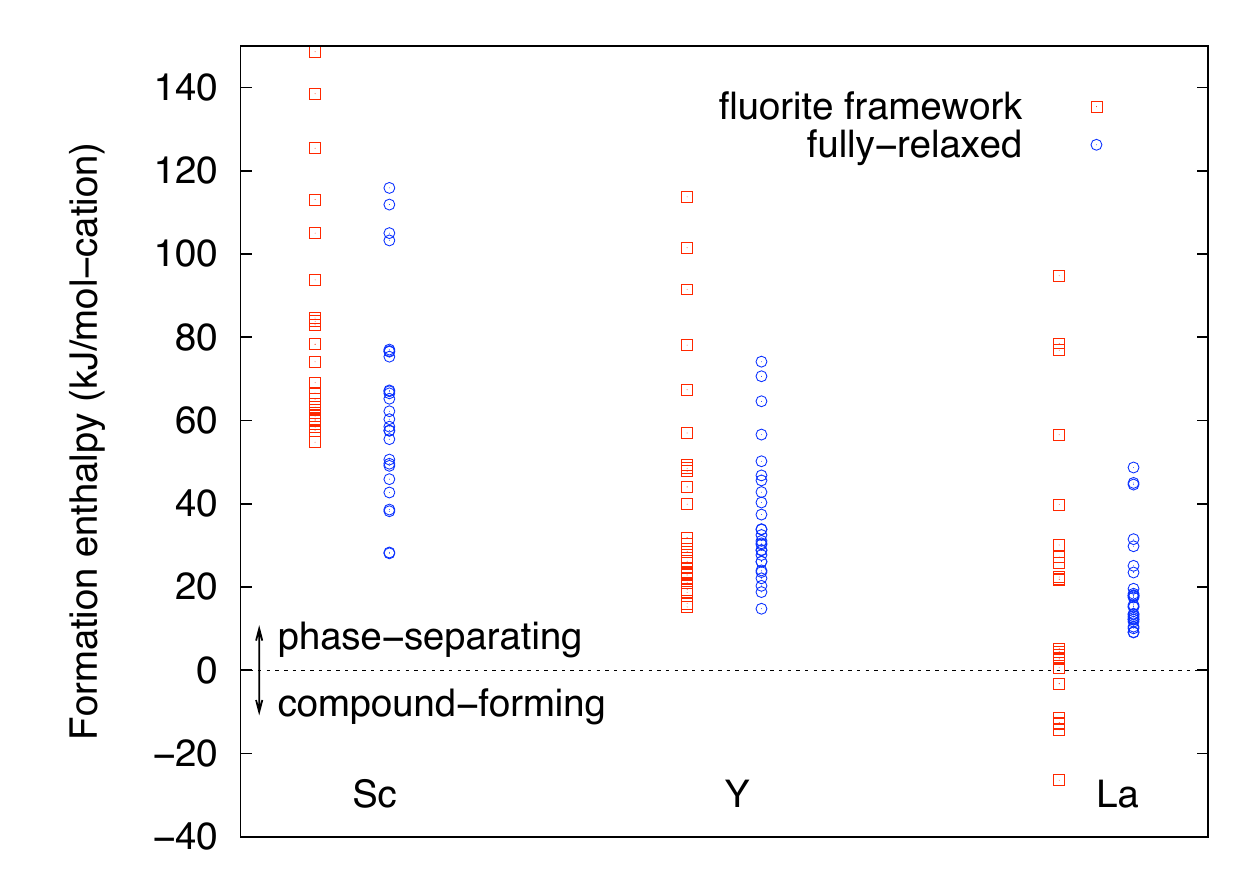}\\
\caption{\label{fig3}The same as in Fig. \ref{fig2} for Th$_{2}$M$_{2}$O$_{7}$ (M = Sc,  Y and La) solid solutions corresponding to $x$=0.5 doping level.}
\end{figure}

In this section we analyze the energetics of compounds with $x$=33\% and 50\% dopant concentrations.  Both compositions are beyond the dopant levels 
$x_{max}$ where the maximum in the measured formation enthalpies are observed for Y and La dopants;\cite{{Aizenshtein2010}} these concentrations are thus 
expected to be in the regime where the dopant cations and oxygen vacancies interact strongly.  The composition A$_{2}$B$_{2}$O$_{7}$ (50\% dopant level) 
also is associated with the widespread group of pyrochlore-structured crystals. The pyrochlore structure is a derivative of fluorite, in which A and B 
cations are ordered along the $<$101$>$ direction, and oxygen vacancies reside in the tetrahedral interstices between adjacent B cations.

Compared to the dilute concentrations discussed above, the number of distinct ionic configurations are much larger at the higher concentrations. In order to determine the energetically preferred atomic arrangements, we performed an exhaustive structure enumeration for Th$_{4}$M$_{2}$O$_{11}$ 
(33\% dopant level) and Th$_{2}$M$_{2}$O$_{7}$ (50\% dopant level) compositions. The former enumeration yielded 117 symmetry-distinct configurations with 
unit cells containing 4 Th and 2 M cations, 11 oxygen ions and 1 oxygen vacancy, arranged over the sites of the ideal fluorite structure.  Similarly, the 
enumeration at 50\% dopant concentration yielded 27 symmetry-distinct fluorite superstructures containing 2 Th and 2 M cations, 7 O ions and 1 oxygen 
vacancy per unit cell. The calculated energies for each of these structures are plotted in Figs.~\ref{fig2} and \ref{fig3} for $x$=0.33 and 0.5, respectively.
In these plots, results labeled "relaxed" correspond to formation energies computed as the energy difference between a fully relaxed 
Th$_{1-x}$M$_{x}$O$_{2-0.5x}$ superstructure and the concentration-weighted average of the energies of relaxed fluorite-structured ThO$_2$ and M$_2$O$_3$ 
in its experimentally-observed low-temperature crystal structure.  The "unrelaxed" results correspond to the analogous energy difference, but using 
structures that are only relaxed with respect to volume, constraining all of the ions to occupy the sites of an ideal fluorite lattice.

The results in Figs.~\ref{fig2} and \ref{fig3} show a large spread in the DFT-calculated formation enthalpies at each fixed composition. For the relaxed 
structures this spread in $\Delta H^f$ between the lowest and highest energy structures is largest for Sc (85.5 and 87.8 kJ/mol-cation at $x$=0.33 and 0.5, 
respectively) and smallest for La (56.6 and 39.6 kJ/mol-cation at $x$=0.33 and 0.5, respectively). All of the formation energies for the relaxed structures 
are positive in Figs.~\ref{fig2} and \ref{fig3}, indicating a preference for phase separation at low temperatures, consistent with experimental measurements 
for La and Y.\cite{Aizenshtein2010} Further, the calculated values of $\Delta H^f$ decrease in magnitude going from Sc to Y to La.  This trend is 
consistent with the results discussed above for the more dilute compositions, and with measured solubility limits at T=1773 K, which are around 50 mol\% 
for La and 20 mol\% for Y, \cite{Aizenshtein2010} and negligible in the case of Sc (the solubility of Sc has not been measured to the best of our knowledge, 
but is expected to be much smaller than for La and Y). 

\begin{figure}
\includegraphics[scale=0.7]{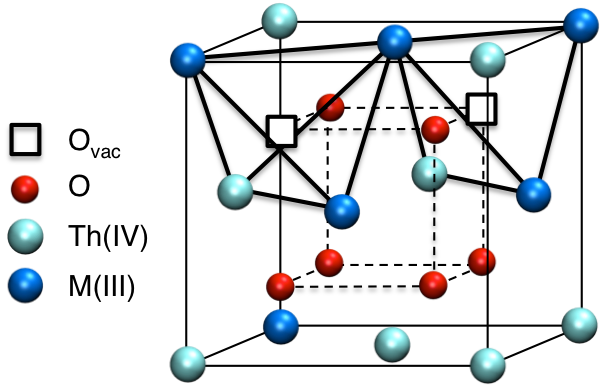}\\
\caption{\label{fig7}Illustration of the stable defect ordering pattern in Th$_{1-x}$M$_{x}$O$_{2-0.5x}$ solid solutions.  The structure is characterized by 
charge-neutral tetrahedra composed of one trivalent dopant M, two corner-shared M and one Th ion with the oxygen vacancies aligned as second nearest neighbors 
along the $<$110$>$ direction.}
\end{figure}

\begin{table}
\caption{\label{table3}Formation enthalpies of Th$_{1-x}$M$_{x}$O$_{2-0.5x}$ (M = Sc, Y, La) structures are listed for the lowest-energy fully-relaxed structures enumerated
in this study.  The structures listed have formation energies within 2 kJ/mol-cation of the lowest energy structure for each dopant at each composition 
considered.  The third column lists the direction of ordering of oxygen-vacancy chains.  The last column lists the number of cations of host (Th) and dopant
(M) species in the 4 tetrahedral nearest-neighbor positions surrounding an oxygen vacancy.}
\begin{ruledtabular}
\begin{tabular}{ccccc}
            & & & & Nearest Neighbors \\
             $x$ & M & $\Delta H^f$ & Direction & Th, M \\
            \hline
           0.5 & Sc & 28.1 & $<$110$>$ & 1, 3 \\
                 & & 28.3 & $<$110$>$ & 3, 1 \\
             & Y  & 14.8 & $<$110$>$ & 1, 3 \\
            & La & 9.1   & $<$110$>$ & 1, 3 \\
             &    & 9.2   & $<$110$>$ & 2, 2 \\
              &   & 10.1 & $<$110$>$ & 1, 3 \\
              &   & 10.5 & $<$110$>$ & 2, 2 \\
                 \hline
          0.33 &  Sc & 22.9 & $<$110$>$ & 1, 3 \\
               &  & 22.9 & $<$110$>$ & 3, 1 \\
               &  & 23.4 & $<$110$>$ & 1, 3 \\
             & Y & 11.7 & $<$110$>$ & 1, 3 \\
              &  & 13.0 & $<$110$>$ & 1, 3 \\
              &  & 13.2 & $<$110$>$ & 1, 3 \\
              &  & 13.5 & $<$110$>$ & 2, 2 \\
              &  &  13.6 & $<$110$>$ & 1, 3 \\
            & La &  4.4  & $<$112$>$ & 4, 0 \\
             &   & 6.1 & $<$112$>$ & 2, 2 \\
             &   & 6.2 & $<$112$>$ & 3, 1 \\
        \end{tabular}
\end{ruledtabular}
\end{table}

The differences in $\Delta H^f$ between the relaxed and unrelaxed structures plotted in Figs.~\ref{fig2} and \ref{fig3} reflect a balance between two competing 
effects: (i) the stabilization of the Th$_{1-x}$M$_x$O$_2$ compounds by atomic relaxation, and (ii) the stabilization of the end-member M$_2$O$_3$ compound when 
it is allowed to change from an ideal fluorite lattice to its experimentally observed low-temperature crystal structure. The first and second effects give rise 
to contributions that decrease and increase the value of $\Delta H^f$, respectively. For the Y dopant, with a size that is most closely matched to Th, the first 
and second effects are comparable in magnitude for the low-energy structures at each composition, such that the formation energies are nearly identical for 
relaxed and unrelaxed structures; the first effect is considerably larger than the second for the high-energy structures, leading to a decrease in the magnitude of $\Delta H^f$ for the relaxed versus unrelaxed structures. For the under-sized Sc dopant, relaxation has the effect of lowering the formation energy for all structures 
considered, while for the over-sized La ion the low-energy and high-energy structures are respectively destabilized and stabilized by atomic relaxations. It is 
interesting to note that in a recent analysis of measured formation energies for La-doped ThO$_2$ solid solutions,\cite{Aizenshtein2010} the mixing energy, 
defined as the formation energy of the solid-solution relative to fluorite-structured ThO$_2$ and {\em fluorite-structured} La$_2$O$_3$, was computed to be 
negative, even though the measured values of $\Delta H^f$ relative to fluorite-structure ThO$_2$ and A-type La$_2$O$_3$, are positive.\cite{Aizenshtein2010}
These results are qualitatively consistent with the present computed energies for La-doped ThO$_2$, which show a change in the sign of the formation energy for 
relaxed versus unrelaxed low-energy structures, driven by the energy difference between fluorite and A-type La$_2$O$_3$ structures.

We consider next the qualitative features of the low and high-energy ionic configurations enumerated in this study. Table \ref{table3} summarizes the formation 
enthalpies of the most stable Th$_{4}$M$_{2}$O$_{11}$ and Th$_{2}$M$_{2}$O$_{7}$ compounds, along with their main structural features, including the direction 
connecting oxygen-vacancy neighbors, and the number of cations of each type (Th and M) in the nearest-neighbor tetrahedron surrounding the vacancy. We 
include such information for several of the lowest-energy structures for each dopant; some of the structures have nearly degenerate or close energies and therefore a number of defect 
ordering patterns may be observed particularly in the presence of entropic effects at elevated temperatures.

We observe that the oxygen vacancies prefer to align as second nearest neighbors along the $<$110$>$ directions for all dopants with an exception of La for $x$=0.33 where the vacancies are ordered along the $<$112$>$ direction. The ordering of oxygen vacancies along the $<$110$>$ direction is also observed for 
the La system at $x$=0.33, but at slightly higher energies (with $\Delta H^f$ $\geq$ 6.7 kJ/mol-cation). An analysis of nearest neighbors around the oxygen 
vacancies shows that the only case in which the oxygen vacancy is surrounded completely by four host Th cations is for the lowest-energy La-doped ThO$_{2}$ 
structure at $x$=0.33. In all other cases we see that the tetrahedra of cations around the vacancy, as pictorially shown in Fig.~\ref{fig7}, are composed of 
both host Th and dopant M ions, with a slight overall preference for binding to trivalent dopant species.

\begin{figure}
\includegraphics[scale=0.7]{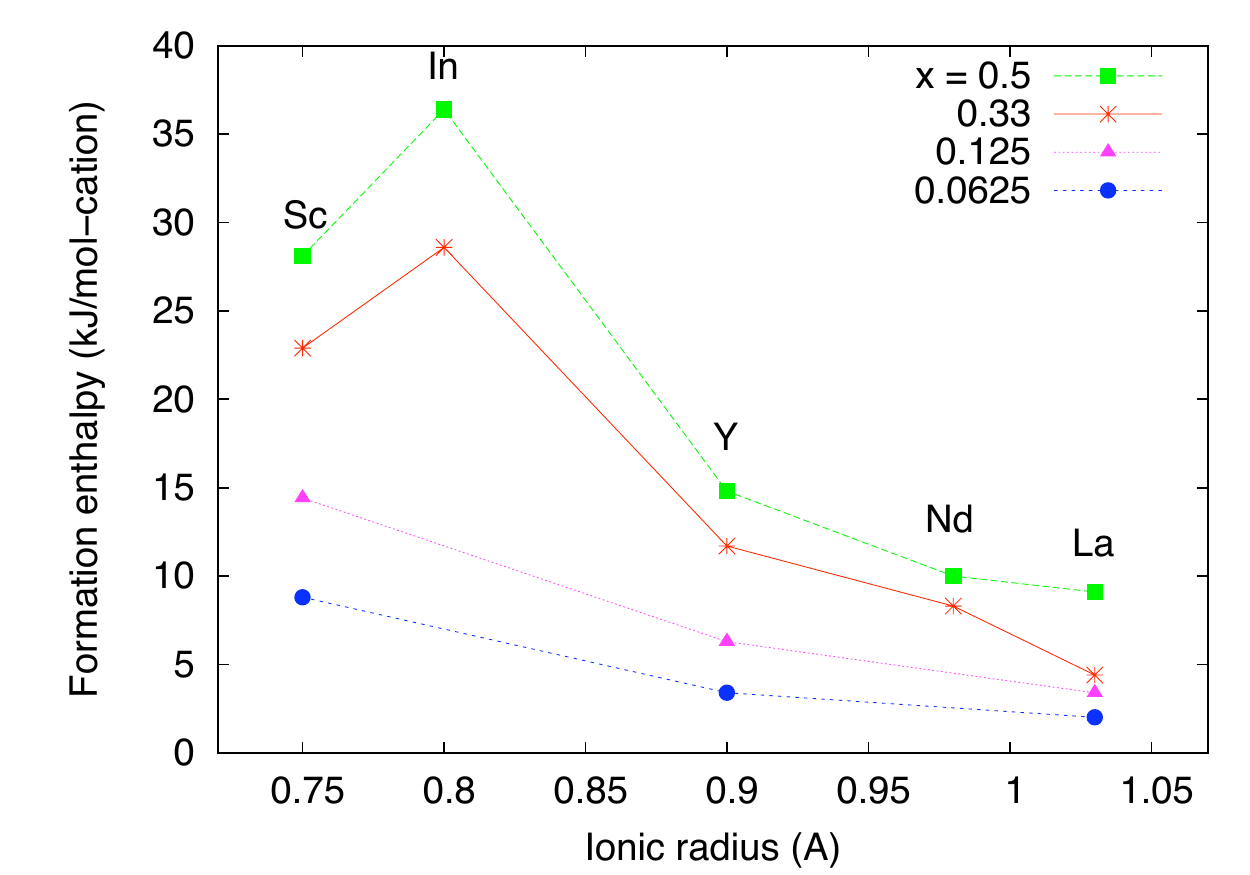}\\
\caption{\label{fig4}Formation enthalpies (kJ/mol-cation) of Th$_{1-x}$M$_{x}$O$_{2-0.5x}$ (M = Sc, In, Y, Nd, La) 
solid solutions at different doping levels $x$ as a function of ionic radius (\AA).}
\end{figure}

For Y dopants at $x$=0.33 and 0.5, the lowest-energy structures are shown in a $<$110$>$ projection in Fig.~\ref{fig6}. In this projection, the smaller (red) 
circles denote $<$110$>$ rows of oxygen ions containing no vacancies, while the darker and lighter larger (blue) circles denote $<$110$>$ columns of Y and Th 
cations, respectively; the black open boxes in the oxygen rows correspond to $<$110$>$ rows of oxygen vacancies.  Both structures in Fig.~\ref{fig6} contain 
the same structural motif, which, in the projected image involves a column of oxygen vacancies located between a triangle of two Y and one Th ion.  This motif 
is illustrated in a different representation in Fig.~\ref{fig7} where it is more clearly apparent that the lowest-energy Y-doped ThO$_2$ structures are 
characterized by a clustering of vacancies as second neighbors on the anion sublattice (i.e., along $<$110$>$) directions, and a clear association of the 
oxygen vacancies with nearest-neighbor dopant cations. In Fig.~\ref{fig7} the motif of the defect cluster involves tetrahedra consisting of three Y and one Th 
cation, surrounding an oxygen vacancy; since these tetrahedra are arranged in a corner-sharing configuration the defect clusters are net charge neutral.  For 
the Th$_{2}$Y$_{2}$O$_{7}$ structure ($x$=0.5) the lowest-energy structure shown in Fig.~\ref{fig6} (left) can also be identified as being characteristic of 
the pyroclore structure with the oxygen vacancies residing in the tetrahedral interstice between adjacent Y dopants, aligned along the $<$110$>$ direction. 

\begin{figure}
\includegraphics[scale=0.7]{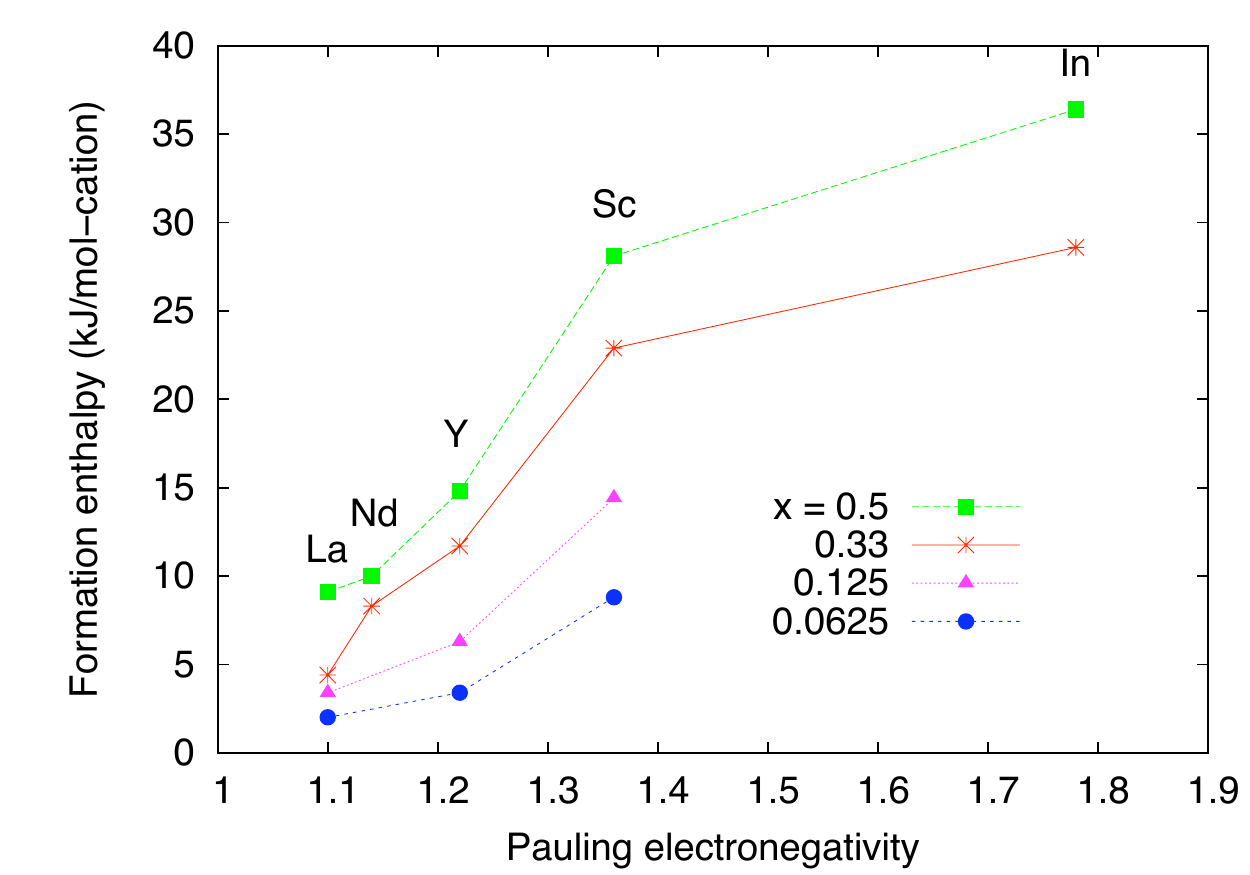}\\
\caption{\label{fig5}Formation enthalpies (kJ/mol-cation) of Th$_{1-x}$M$_{x}$O$_{2-0.5x}$ (M = Sc, In, Y, Nd, La) 
solid solutions at different doping levels $x$ as a function of Pauling electronegativities.}
\end{figure}

\begin{figure*}
\includegraphics[scale=0.8]{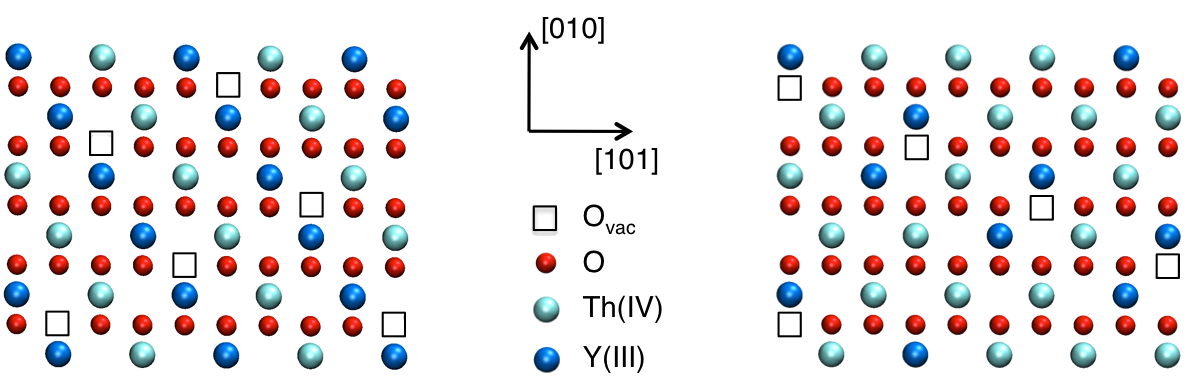}\\
\caption{\label{fig6}Lowest-energy atomic arrangements in Th$_{1-x}$Y$_{x}$O$_{2-0.5x}$ solid solutions at $x$=0.5 (left) and 0.33 (right). The different circle symbols correspond to rows of O, Th and Y ions directed along the $<$110$>$ direction (out of the plane of the page), while the open boxes correspond to rows of oxygen vacancies along the same direction. For clarity all atoms are displayed in their ideal fluorite positions.}
\end{figure*}

\begin{figure}
\includegraphics[scale=0.5]{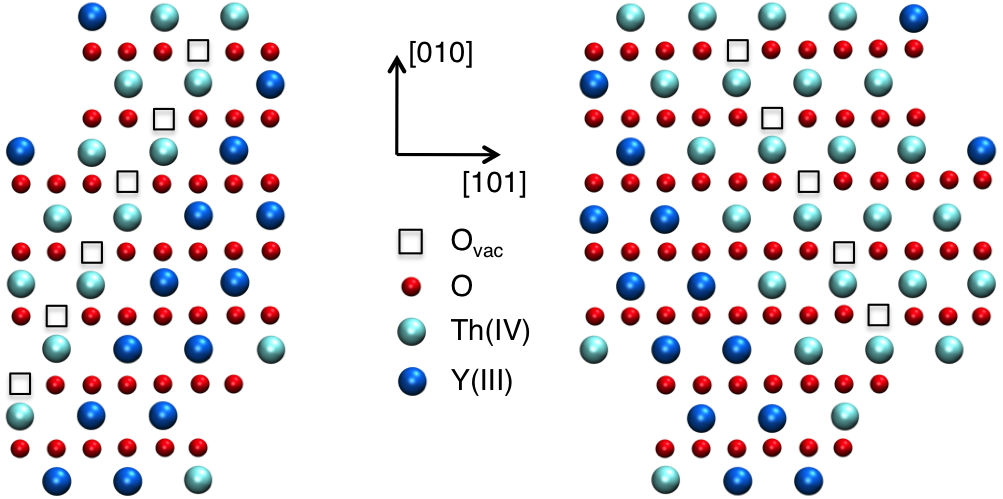}\\
\caption{\label{fig8}Highest-energy atomic arrangements in Th$_{1-x}$M$_{x}$O$_{2-0.5x}$ solid solutions at $x$=0.5 (left) and 0.33 (right). The symbols have the same meanining as in Fig. \ref{fig6}. For clarity 
all atoms are displayed in their perfect fluorite positions.}
\end{figure}

The structures of the highest-energy configurations for Y-doped ThO$_2$ considered in our calculations are illustrated in Fig.~\ref{fig8}.  These structures 
are characterized by the location of rows of oxygen vacancies between host Th ions, and thus contain large regions that are locally charge imbalanced. The 
large energy difference between these configurations and those characteristic of the lowest-energy structures demonstrates a strong preference for association 
of oxygen-vacancy chains with dopant cation species in this system.

\section{Discussion}

The results in Figs. ~\ref{fig2} and ~\ref{fig3} and Tables \ref{table2} and \ref{table3} feature a strong variation in the calculated values of $\Delta H^f$ 
as the dopant species change from Sc to Y to La.  To gain further insight to the origins of this trend, we explore the correlations between $\Delta H^f$ and 
the dopant properties listed in Table \ref{table1}.  Figure \ref{fig4} plots calculated formation energies versus ionic radius for the lowest energy structures 
containing Sc, Y, and La dopants, as well as the two additional trivalent dopants In and Nd. A tendency towards decreasing magnitudes of formation energies for dopants with 
increasing size is observed, but with the notable exception of In which has the largest positive calculated value of $\Delta H^f$ and an ionic radius larger than Sc.  

An interesting correlation is shown in Fig.~\ref{fig5}, which plots $\Delta H^f$ versus the Pauling electronegativity ($\chi$) of the dopant species.  A 
monotonic decrease is observed in the calculated values of $\Delta H^f$ with decreasing $\chi$.  Since a decrease in the electronegativity of the dopant cation 
corresponds to an increasing electronegativity {\em difference} between it and oxygen, the trend in Fig.~\ref{fig5} suggests that a decreasing ionic character 
in the bonding between dopant and oxygen ions has a tendency to destabilize doped ThO$_2$ mixtures.

To investigate this correlation further, we performed an analysis of the electronic structures of the lowest-energy enumerated compounds, through calculations 
of Bader charges and partial electronic densities of states (DOS). For low-energy Sc and La doped structures, the calculated Bader charge on La is larger than 
that on Sc by appoximately 0.1 electrons, consistent with the higher value of $\chi$ for the latter. The change in the ionic character of the bonding between 
oxygen and dopant cations with varying $\chi$ is also indicated in the calculated partial DOS. The calculated DOS feature a partial covalent character to the 
bonding, characterized by a weak hybridization of O $2p$ and dopant $d$ electrons.  The degree of hybridization, as measured from the integrated number of 
electrons in the O $2p$ band, is found to decrease monotonically going from Sc to In to Y to Nd to La, consistent with the trend in the dopant 
electronegativities. Overall, the analysis suggests that the mixing energetics of aliovalently-doped ThO$_{2}$ solid-solutions is influenced not only by the 
size and charge of the dopant species, but also by chemical effects associated with the degree of covalency characterizing the bonding between dopant and oxygen 
ions.

It is interesting to compare the results of the present study with those obtained in previous investigations of related trivalent-doped fluorite-structured solid solutions.
In the most stable $\delta$-phase of YSZ each oxygen vacancy is enclosed in a tetrahedron consisting of two Y and two Zr cations, and the vacancies are ordered 
along the $<$111$>$ direction, as third-nearest neighbors. This atomic arrangement exhibits a negative formation enthalpy, indicating a compound-forming tendency. 
On the contrary, in trivalently-doped ThO$_2$ we observe only positive enthalpies of formation for all dopants considered, and an overall preference for $<$110$>$ 
ordering of oxygen vacancies (i.e.,  as second-nearest neighbors on the anion simple-cubic sublattice).  In agreement with the present results, it has been observed 
that formation enthalpies of aliovalently-doped ZrO$_2$ are most positive for small dopants.\cite{Bogicevic2001a} The same trend has been found for 
pyrochlore-structured solid solutions of CeO$_2$ with trivalent dopants,\cite{Minervini1999} where the formation energy increases in magnitude going from La to Gd to Y dopants. 
It also has been demonstrated in the dilute regime that small dopants exhibit the strongest binding with the neighboring oxygen vacancies in CeO$_2$, which 
is again consistent with the results of the present study.

The variations in binding and formation energies found for the different dopants considered in this study may have important consequences for oxygen ionic 
conductivity in aliovalently-doped ThO$_2$ compounds.  Specifically, La dopants are found to have significantly smaller formation energies and binding energies 
(see Table \ref{table2}) than the other dopants considered in this work.  Further, we found that the La-doped compounds showed a higher degeneracy in the energy 
of competing structures at more concentrated compositions.  Overall, these trends are consistent with experimental observations. Specifically, the solubility of 
La in ThO$_2$ is significantly larger that that for Y or Sc doped systems, and from an analysis of the calorimetry results in Ref.\cite{Aizenshtein2010} the 
association energy for defects in La-doped ThO$_2$ is estimated to be considerably smaller than in Y-doped ThO$_2$. The smaller magnitudes for the formatin energies and oxygen-vacancy binding energies for La dopants is expected to give rise to higher ionic conductivities in this system, relative to Y-doped ThO$_2$.  While 
conductivity data for the latter system are available,\cite{Lasker1966} measurements for the former have not yet been published, to the best of our knowledge.
Such measurements for La-doped ThO$_2$ would thus be interesting to pursue in light of the present results.

\section{Conclusions}

In this study we have employed DFT calculations to study formation energetics and defect-ordering tendencies in ThO$_2$ compounds doped with the trivalent 
cation species:  Sc, In, Y, Nd and La.  The main results derived from these calculations can be summarized as follows:

(1) Solid solutions of fluorite-structured ThO$_{2}$ with oxides of trivalent cations (Sc, In, Y, Nd, La) exhibit positive enthalpies of formation from constituent 
oxides, indicative of a phase-separating tendency across all dopant concentrations considered (6.25, 12.5, 33.3 and 50\%). This strongly suggests that there are no underlying stable long-range ordered phases in these compounds.  However, the large variation in energy for different dopant/vacancy arrangements suggests an appreciable driving force for short-ranged ordering (clustering) in these solid solutions at finite temperatures. 

(2) For a given dopant concentration the formation enthalpies tend to decrease in magnitude as the size and electronegativity of the trivalent dopant decrease.  This 
trend is consistent with available experimental data on solubility limits which are highest for Nd- and La-doped ThO$_{2}$. 

(3) The lowest-energy atomic structures are characterized by a repulsion of oxygen vacancies as nearest neighbors.  Oxygen vacancies are found to prefer to 
align as second nearest neighbors along the $<$110$>$ direction for all of the dopants and concentrations considered, with the exception of 33.3\% La-doped 
ThO$_{2}$ where they order as more distant neighbors along the $<$112$>$ direction.

(4) The solute-vacancy binding energies in ThO$_{2}$ solid solutions with under-sized Sc and intermediate-sized Y dopants are estimated to be on the order of 
a few kJ/mol-cation.  For over-sized La dopants, this binding energy is estimated to be significantly smaller in magnitude. 

(5) The small magnitude of the formation and binding enthalpies for La-doped ThO$_{2}$ solid solutions is expected to result in an ionic conductivity that is 
higher in this system than for the other trivalent-doped systems considered in this study.

\section{Ackowledgements}
This work was supported as part of the Materials Science of Actinides, an Energy Frontier Research Center funded by the U.S. Department of Energy, 
Office of Science, Office of Basic Energy Sciences under award number DE-SC0001089.  The authors acknowledge fruitful discussions with Dr. 
Tatiana Shvareva.  We also acknowledge very helpful input from Dr. Axel van de Walle on the application of the ATAT package in this work.

\newpage


\begin{thebibliography}{28}
\expandafter\ifx\csname natexlab\endcsname\relax\def\natexlab#1{#1}\fi
\expandafter\ifx\csname bibnamefont\endcsname\relax
  \def\bibnamefont#1{#1}\fi
\expandafter\ifx\csname bibfnamefont\endcsname\relax
  \def\bibfnamefont#1{#1}\fi
\expandafter\ifx\csname citenamefont\endcsname\relax
  \def\citenamefont#1{#1}\fi
\expandafter\ifx\csname url\endcsname\relax
  \def\url#1{\texttt{#1}}\fi
\expandafter\ifx\csname urlprefix\endcsname\relax\def\urlprefix{URL }\fi
\providecommand{\bibinfo}[2]{#2}
\providecommand{\eprint}[2][]{\url{#2}}

\bibitem[{\citenamefont{Minh}(1993)}]{Minh1993}
\bibinfo{author}{\bibfnamefont{N.~Q.} \bibnamefont{Minh}},
  \bibinfo{journal}{Journal of the American Ceramic Society}
  \textbf{\bibinfo{volume}{76}}, \bibinfo{pages}{563} (\bibinfo{year}{1993}).
  
    \bibitem[{\citenamefont{Kharton}(2000)}]{Kharton2000}
  \bibinfo{author}{\bibfnamefont{V.~V.} \bibnamefont{Kharton}},
  \bibinfo{author}{\bibfnamefont{A.~A.} \bibnamefont{Yaremchenko}},
  \bibinfo{author}{\bibfnamefont{E.~N.} \bibnamefont{Naumovich}}, and
  \bibinfo{author}{\bibfnamefont{F.~M.~B.} \bibnamefont{Marques}},
  \bibinfo{journal}{Journal of Solid State Electrochemistry}
  \textbf{\bibinfo{volume}{4}}, \bibinfo{pages}{243} (\bibinfo{year}{2000}).
  
  \bibitem[{\citenamefont{Steele}(2001)}]{Steele2001}
  \bibinfo{author}{\bibfnamefont{B.~C.~H.} \bibnamefont{Steele}} and
  \bibinfo{author}{\bibfnamefont{A.} \bibnamefont{Heinzel}},
  \bibinfo{journal}{Nature}
  \textbf{\bibinfo{volume}{414}}, \bibinfo{pages}{345} (\bibinfo{year}{2001}).
  
    \bibitem[{\citenamefont{Jiang}(2004)}]{Jiang2004}
  \bibinfo{author}{\bibfnamefont{S.~P.} \bibnamefont{Jiang}} and
  \bibinfo{author}{\bibfnamefont{S.~H.} \bibnamefont{Chan}},
  \bibinfo{journal}{Journal of Materials Science}
  \textbf{\bibinfo{volume}{39}}, \bibinfo{pages}{4405} (\bibinfo{year}{2004}).
  
    \bibitem[{\citenamefont{Kharton}(2008)}]{Kharton2008}
  \bibinfo{author}{\bibfnamefont{E.~V.} \bibnamefont{Tsipis}} and
  \bibinfo{author}{\bibfnamefont{V.~V.} \bibnamefont{Kharton}},
  \bibinfo{journal}{Journal of Solid State Electrochemistry}
  \textbf{\bibinfo{volume}{12}}, \bibinfo{pages}{1367} (\bibinfo{year}{2008}).
  
     \bibitem[{\citenamefont{Trubelja}(1991)}]{Trubelja1991}
  \bibinfo{author}{\bibfnamefont{M.~F.} \bibnamefont{Trubelja}} and
  \bibinfo{author}{\bibfnamefont{V.~S.} \bibnamefont{Stubician}},
  \bibinfo{journal}{Journal of the American Ceramic Society}
  \textbf{\bibinfo{volume}{74}}, \bibinfo{pages}{2489} (\bibinfo{year}{1991}).
  
     \bibitem[{\citenamefont{Goff}(1999)}]{Goff1999}
  \bibinfo{author}{\bibfnamefont{J.~P.} \bibnamefont{Goff}},
  \bibinfo{author}{\bibfnamefont{W.} \bibnamefont{Hayes}},
  \bibinfo{author}{\bibfnamefont{S.} \bibnamefont{Hull}},
    \bibinfo{author}{\bibfnamefont{M.~T.} \bibnamefont{Hutchings}}, and
      \bibinfo{author}{\bibfnamefont{K.~N.} \bibnamefont{Clausen}},
  \bibinfo{journal}{Physical Review B}
  \textbf{\bibinfo{volume}{59}}, \bibinfo{pages}{14202} (\bibinfo{year}{1999}).
  
      \bibitem[{\citenamefont{Chen}(2006)}]{Chen2006}
  \bibinfo{author}{\bibfnamefont{W.~Q.} \bibnamefont{Chen}} and
  \bibinfo{author}{\bibfnamefont{A.} \bibnamefont{Navrotksy}},
  \bibinfo{journal}{Journal of Materials Research}
  \textbf{\bibinfo{volume}{21}}, \bibinfo{pages}{3242} (\bibinfo{year}{2006}).
  
      \bibitem[{\citenamefont{Navrotsky}(2007)}]{Navrotsky2007}
  \bibinfo{author}{\bibfnamefont{A.} \bibnamefont{Navrotsky}},
  \bibinfo{author}{\bibfnamefont{P.} \bibnamefont{Simoncic}},
  \bibinfo{author}{\bibfnamefont{H.} \bibnamefont{Yokokawa}},
  \bibinfo{author}{\bibfnamefont{W.~Q.} \bibnamefont{Chen}}, and
  \bibinfo{author}{\bibfnamefont{T.} \bibnamefont{Lee}},
  \bibinfo{journal}{Faraday Discussions}
  \textbf{\bibinfo{volume}{134}}, \bibinfo{pages}{171} (\bibinfo{year}{2007}).
  
     \bibitem[{\citenamefont{Avila-Paredes}(2009)}]{Avila-Paredes2009}
  \bibinfo{author}{\bibfnamefont{H.~J.} \bibnamefont{Avila-Paredes}},
  \bibinfo{author}{\bibfnamefont{T.} \bibnamefont{Shvareva}},
  \bibinfo{author}{\bibfnamefont{W.} \bibnamefont{Chen}},
  \bibinfo{author}{\bibfnamefont{A.} \bibnamefont{Navrotsky}}, and
  \bibinfo{author}{\bibfnamefont{S.} \bibnamefont{Kim}},
  \bibinfo{journal}{Physical Chemistry Chemical Physics}
  \textbf{\bibinfo{volume}{11}}, \bibinfo{pages}{8580} (\bibinfo{year}{2009}).
      
        \bibitem[{\citenamefont{Aizenshtein}(2010)}]{Aizenshtein2010}
  \bibinfo{author}{\bibfnamefont{M.} \bibnamefont{Aizenshtein}},
  \bibinfo{author}{\bibfnamefont{T.} \bibnamefont{Shvareva}}, and
  \bibinfo{author}{\bibfnamefont{A.} \bibnamefont{Navrotsky}},
  \bibinfo{journal}{Journal of the American Ceramic Society, accepted.}
  
    \bibitem[{\citenamefont{Lee}(2003)}]{Lee2003}
  \bibinfo{author}{\bibfnamefont{T.~A.} \bibnamefont{Lee}},
  \bibinfo{author}{\bibfnamefont{A.} \bibnamefont{Navrotksy}}, and
  \bibinfo{author}{\bibfnamefont{I.} \bibnamefont{Molodetsky}},
  \bibinfo{journal}{Journal of Materials Research}
  \textbf{\bibinfo{volume}{18}}, \bibinfo{pages}{908} (\bibinfo{year}{2003}).
    
    \bibitem[{\citenamefont{Tien}(1963)}]{Tien1963}
  \bibinfo{author}{\bibfnamefont{T.~Y.} \bibnamefont{Tien}} and
  \bibinfo{author}{\bibfnamefont{E.~S.} \bibnamefont{Subbarao}},
  \bibinfo{journal}{Journal of Chemical Physics}
  \textbf{\bibinfo{volume}{39}}, \bibinfo{pages}{1041} (\bibinfo{year}{1963}).
  
  
       \bibitem[{\citenamefont{Bogicevic}(2001)}]{Bogicevic2001a}
  \bibinfo{author}{\bibfnamefont{A.} \bibnamefont{Bogicevic}},
  \bibinfo{author}{\bibfnamefont{C.} \bibnamefont{Wolverton}},
   \bibinfo{author}{\bibfnamefont{G.~M.} \bibnamefont{Crosbie}}, and
    \bibinfo{author}{\bibfnamefont{E.~B.} \bibnamefont{Stechel}},
  \bibinfo{journal}{Physical Review B}
  \textbf{\bibinfo{volume}{64}}, \bibinfo{pages}{014106} (\bibinfo{year}{2001}).
  
    \bibitem[{\citenamefont{Bogicevic}(2001)}]{Bogicevic2001}
  \bibinfo{author}{\bibfnamefont{A.} \bibnamefont{Bogicevic}} and
  \bibinfo{author}{\bibfnamefont{C.} \bibnamefont{Wolverton}},
  \bibinfo{journal}{Europhysics Letters}
  \textbf{\bibinfo{volume}{56}}, \bibinfo{pages}{393} (\bibinfo{year}{2001}).
  
    \bibitem[{\citenamefont{Khan}(1998)}]{Khan1998}
  \bibinfo{author}{\bibfnamefont{M.~S.} \bibnamefont{Khan}},
  \bibinfo{author}{\bibfnamefont{M.~S.} \bibnamefont{Islam}}, and
  \bibinfo{author}{\bibfnamefont{D.~R.} \bibnamefont{Bates}},
  \bibinfo{journal}{Journal of Materials Chemistry}
  \textbf{\bibinfo{volume}{8}}, \bibinfo{pages}{2299} (\bibinfo{year}{1998}).
  
    \bibitem[{\citenamefont{Zacate}(2000)}]{Zacate2000}
  \bibinfo{author}{\bibfnamefont{M.~O.} \bibnamefont{Zacate}},
  \bibinfo{author}{\bibfnamefont{L.} \bibnamefont{Minervini}},
  \bibinfo{author}{\bibfnamefont{D.~J.} \bibnamefont{Bradfield}},
  \bibinfo{author}{\bibfnamefont{R~.W.} \bibnamefont{Grimes}}, and
  \bibinfo{author}{\bibfnamefont{K.~E.} \bibnamefont{Sickafus}},
  \bibinfo{journal}{Solid State Ionics}
  \textbf{\bibinfo{volume}{128}}, \bibinfo{pages}{243} (\bibinfo{year}{2000}).
  
        \bibitem[{\citenamefont{Predith}(2008)}]{Predith2008}
  \bibinfo{author}{\bibfnamefont{A.} \bibnamefont{Predith}},
  \bibinfo{author}{\bibfnamefont{G.} \bibnamefont{Ceder}},
  \bibinfo{author}{\bibfnamefont{C.} \bibnamefont{Wolverton}},
  \bibinfo{author}{\bibfnamefont{K.} \bibnamefont{Persson}}, and
  \bibinfo{author}{\bibfnamefont{T.} \bibnamefont{Mueller}},
  \bibinfo{journal}{Physical Review B}
  \textbf{\bibinfo{volume}{77}}, \bibinfo{pages}{144104} (\bibinfo{year}{2008}).
  
        \bibitem[{\citenamefont{Xia}(2009)}]{Xia2009}
  \bibinfo{author}{\bibfnamefont{X.} \bibnamefont{Xia}},
  \bibinfo{author}{\bibfnamefont{R.} \bibnamefont{Oldman}}, and
  \bibinfo{author}{\bibfnamefont{R.} \bibnamefont{Catlow}},
  \bibinfo{journal}{Chemistry of Materials}
  \textbf{\bibinfo{volume}{21}}, \bibinfo{pages}{3576} (\bibinfo{year}{2009}).
  
          \bibitem[{\citenamefont{Ray}(1980)}]{Ray1980}
  \bibinfo{author}{\bibfnamefont{S.~P.} \bibnamefont{Ray}},
  \bibinfo{author}{\bibfnamefont{V.~S.} \bibnamefont{Stubican}}, and
  \bibinfo{author}{\bibfnamefont{D.~E.} \bibnamefont{Cox}},
  \bibinfo{journal}{Materials Research Bulletin}
  \textbf{\bibinfo{volume}{15}}, \bibinfo{pages}{1419} (\bibinfo{year}{1980}).
  
  
      \bibitem[{\citenamefont{Pryde}(1995)}]{Pryde1995}
  \bibinfo{author}{\bibfnamefont{A.~K.~A.} \bibnamefont{Pryde}},
  \bibinfo{author}{\bibfnamefont{S.} \bibnamefont{Vyas}},
  \bibinfo{author}{\bibfnamefont{R.~W.} \bibnamefont{Grimes}},
  \bibinfo{author}{\bibfnamefont{J.~A.} \bibnamefont{Gardner}}, and
  \bibinfo{author}{\bibfnamefont{R.} \bibnamefont{Wang}},
  \bibinfo{journal}{Physical Review B}
  \textbf{\bibinfo{volume}{52}}, \bibinfo{pages}{13214} (\bibinfo{year}{1995}).
  
      \bibitem[{\citenamefont{Minervini}(1999)}]{Minervini1999}
  \bibinfo{author}{\bibfnamefont{L.} \bibnamefont{Minervini}},
  \bibinfo{author}{\bibfnamefont{M.~O.} \bibnamefont{Zacate}}, and
  \bibinfo{author}{\bibfnamefont{R.~W.} \bibnamefont{Grimes}},
  \bibinfo{journal}{Solid State Ionics}
  \textbf{\bibinfo{volume}{116}}, \bibinfo{pages}{339} (\bibinfo{year}{1999}).  
  
      \bibitem[{\citenamefont{Lung}(1998)}]{Lung1998}
  \bibinfo{author}{\bibfnamefont{M.} \bibnamefont{Lung}} and
  \bibinfo{author}{\bibfnamefont{O.} \bibnamefont{Gremm}},
  \bibinfo{journal}{Nuclear Engineering and Design}
  \textbf{\bibinfo{volume}{180}}, \bibinfo{pages}{133} (\bibinfo{year}{1998}).
  
      \bibitem[{\citenamefont{Lombardi}(2008)}]{Lombardi2008}
  \bibinfo{author}{\bibfnamefont{C.} \bibnamefont{Lombardi}},
  \bibinfo{author}{\bibfnamefont{L.} \bibnamefont{Luzzi}},
  \bibinfo{author}{\bibfnamefont{E.} \bibnamefont{Padovani}}, and
  \bibinfo{author}{\bibfnamefont{F.} \bibnamefont{Vettraino}},
  \bibinfo{journal}{Progress in Nuclear Energy}
  \textbf{\bibinfo{volume}{50}}, \bibinfo{pages}{944} (\bibinfo{year}{2008}).

  

     \bibitem[{\citenamefont{Stefanovich}(1994)}]{Stefanovich1994}
  \bibinfo{author}{\bibfnamefont{E.~V.} \bibnamefont{Stefanovich}},
  \bibinfo{author}{\bibfnamefont{A.~L.} \bibnamefont{Shluger}}, and
  \bibinfo{author}{\bibfnamefont{C.~R.~A.} \bibnamefont{Catlow}},
  \bibinfo{journal}{Physical Review B}
  \textbf{\bibinfo{volume}{49}}, \bibinfo{pages}{11560} (\bibinfo{year}{1994}).

    \bibitem[{\citenamefont{Stapper}(1999)}]{Stapper1999}
  \bibinfo{author}{\bibfnamefont{G.} \bibnamefont{Stapper}},
  \bibinfo{author}{\bibfnamefont{M.} \bibnamefont{Bernasconi}},
  \bibinfo{author}{\bibfnamefont{N.} \bibnamefont{Nicoloso}}, and
  \bibinfo{author}{\bibfnamefont{M.} \bibnamefont{Parrinello}},
  \bibinfo{journal}{Physical Review B}
  \textbf{\bibinfo{volume}{59}}, \bibinfo{pages}{797} (\bibinfo{year}{1999}).

     \bibitem[{\citenamefont{Shannon}(1976)}]{Shannon1976}
  \bibinfo{author}{\bibfnamefont{R.~D.} \bibnamefont{Shannon}},
  \bibinfo{journal}{Acta Crystallographica A}
  \textbf{\bibinfo{volume}{32}}, \bibinfo{pages}{751} (\bibinfo{year}{1976}).

     \bibitem[{\citenamefont{PBE}(1996)}]{Perdew1996}
  \bibinfo{author}{\bibfnamefont{J.~P.} \bibnamefont{Perdew}},
  \bibinfo{author}{\bibfnamefont{K.} \bibnamefont{Burke}}, and
  \bibinfo{author}{\bibfnamefont{M.} \bibnamefont{Ernzerhof}},
  \bibinfo{journal}{Physical Review Letters}
  \textbf{\bibinfo{volume}{77}}, \bibinfo{pages}{3865} (\bibinfo{year}{1996}).
  
      \bibitem[{\citenamefont{Blochl}(1994)}]{Blochl1994}
  \bibinfo{author}{\bibfnamefont{P.~E.} \bibnamefont{Bl\"{o}chl}},
  \bibinfo{journal}{Physical Review B}
  \textbf{\bibinfo{volume}{50}}, \bibinfo{pages}{17953} (\bibinfo{year}{1994}).
  
       \bibitem[{\citenamefont{Kresse}(1993)}]{Kresse1993}
  \bibinfo{author}{\bibfnamefont{G.} \bibnamefont{Kresse}} and
  \bibinfo{author}{\bibfnamefont{J.} \bibnamefont{Hafner}},
  \bibinfo{journal}{Physical Review B}
  \textbf{\bibinfo{volume}{48}}, \bibinfo{pages}{13115} (\bibinfo{year}{1993}).
  
        \bibitem[{\citenamefont{Kresse}(1993)}]{Kresse1994}
  \bibinfo{author}{\bibfnamefont{G.} \bibnamefont{Kresse}} and
  \bibinfo{author}{\bibfnamefont{J.} \bibnamefont{Hafner}},
  \bibinfo{journal}{Physical Review B}
  \textbf{\bibinfo{volume}{49}}, \bibinfo{pages}{14251} (\bibinfo{year}{1994}).
    
     \bibitem[{\citenamefont{Sevik}(2009)}]{Sevik2009}
  \bibinfo{author}{\bibfnamefont{C.} \bibnamefont{Sevik}} and
  \bibinfo{author}{\bibfnamefont{T.} \bibnamefont{\c{C}a\u{g}in}},
  \bibinfo{journal}{Physical Review B}
  \textbf{\bibinfo{volume}{80}}, \bibinfo{pages}{014108} (\bibinfo{year}{2009}).
  
    \bibitem[{\citenamefont{Dudarev}(1998)}]{Dudarev1998}
  \bibinfo{author}{\bibfnamefont{S.~L.} \bibnamefont{Dudarev}},
  \bibinfo{author}{\bibfnamefont{G.~A.} \bibnamefont{Botton}},
  \bibinfo{author}{\bibfnamefont{S.~Y.} \bibnamefont{Savrasov}},
  \bibinfo{author}{\bibfnamefont{C.~J.} \bibnamefont{Humphreys}}, and
  \bibinfo{author}{\bibfnamefont{A.~P.} \bibnamefont{Sutton}},
  \bibinfo{journal}{Physical Review B}
  \textbf{\bibinfo{volume}{57}}, \bibinfo{pages}{1505} (\bibinfo{year}{1998}).
  
       \bibitem[{\citenamefont{Monkhorst}(1976)}]{Monkhorst1976}
  \bibinfo{author}{\bibfnamefont{H.~J.} \bibnamefont{Monkhorst}} and
  \bibinfo{author}{\bibfnamefont{J.~D.} \bibnamefont{Pack}},
  \bibinfo{journal}{Physical Review B}
  \textbf{\bibinfo{volume}{13}}, \bibinfo{pages}{5188} (\bibinfo{year}{1976}).
  
      \bibitem[{\citenamefont{Blochl}(1994)}]{Blochl1994a}
  \bibinfo{author}{\bibfnamefont{P.~E.} \bibnamefont{Bl\"{o}chl}},
  \bibinfo{author}{\bibfnamefont{O.} \bibnamefont{Jepsen}}, and
  \bibinfo{author}{\bibfnamefont{O.~K.} \bibnamefont{Andersen}},
  \bibinfo{journal}{Physical Review B}
  \textbf{\bibinfo{volume}{49}}, \bibinfo{pages}{16223} (\bibinfo{year}{1994}).
  
  
  
  
       \bibitem[{\citenamefont{Walle}(2002)}]{Walle2002}
  \bibinfo{author}{\bibfnamefont{A.} \bibnamefont{van de Walle}},
  \bibinfo{author}{\bibfnamefont{M.} \bibnamefont{Asta}}, and
    \bibinfo{author}{\bibfnamefont{G.} \bibnamefont{Ceder}},
  \bibinfo{journal}{CALPHAD}
  \textbf{\bibinfo{volume}{26}}, \bibinfo{pages}{539} (\bibinfo{year}{2002}).
  
       \bibitem[{\citenamefont{Walle}(2009)}]{Walle2009}
  \bibinfo{author}{\bibfnamefont{A.} \bibnamefont{van de Walle}},
  \bibinfo{journal}{CALPHAD}
  \textbf{\bibinfo{volume}{33}}, \bibinfo{pages}{266} (\bibinfo{year}{2009}). 
  
       \bibitem[{\citenamefont{Bugaev}(2002)}]{Bugaev2002}
  \bibinfo{author}{\bibfnamefont{V.~N.} \bibnamefont{Bugaev}},
  \bibinfo{author}{\bibfnamefont{H.} \bibnamefont{Reichert}},
    \bibinfo{author}{\bibfnamefont{O.} \bibnamefont{Shchyglo}},
     \bibinfo{author}{\bibfnamefont{O.} \bibnamefont{Udyansky}},
      \bibinfo{author}{\bibfnamefont{Y.} \bibnamefont{Sikula}}, and
       \bibinfo{author}{\bibfnamefont{H.} \bibnamefont{Dosch}},
  \bibinfo{journal}{Physical Review B}
  \textbf{\bibinfo{volume}{65}}, \bibinfo{pages}{180203} (\bibinfo{year}{2002}).
  
         \bibitem[{\citenamefont{Lasker}(1966)}]{Lasker1966}
  \bibinfo{author}{\bibfnamefont{M.~F.} \bibnamefont{Lasker}} and
  \bibinfo{author}{\bibfnamefont{R.~A.} \bibnamefont{Rapp}},
  \bibinfo{journal}{Zeitschrift f\"{u}r Physikalische Chemie}
  \textbf{\bibinfo{volume}{49}}, \bibinfo{pages}{198} (\bibinfo{year}{1966}).
  
\end{thebibliography}
\end{document}